\documentclass[proceedings]{stacs}
\stacsheading{2009}{433--444}{Freiburg}
\firstpageno{433}

\usepackage[latin1]{inputenc} 
\usepackage[T1]{fontenc}
\usepackage{color}
\usepackage{amssymb}
\usepackage{amsfonts}
\usepackage{amsmath}
\usepackage{stmaryrd}
\usepackage{times}
\usepackage{latexsym}

\newcommand\Eval[1]{\left\llbracket{#1}\right\rrbracket}

\newcommand\uuarrow{\rlap{$\uparrow$}\raise.5ex\hbox{$\uparrow$}}
\newcommand\ddarrow{\rlap{$\downarrow$}\raise.5ex\hbox{$\downarrow$}}
\newcommand\nat{\mathbb{N}}
\newcommand\Z{\mathbb{Z}}
\newcommand\pow{\mathbb{P}}
\newcommand\Sober{\mathcal S}
\newcommand\Hoare{\mathcal H}
\newcommand\LUB{\mathrm{Lub}}
\newcommand\IND{\mathrm{Ind}}
\newcommand\Lim{\mathrm{L}}

\newcommand\dom{\mathop{\mathrm{dom}}}

\newcommand\mopen{\{\mkern-\thinmuskip|}
\newcommand\mclose{|\mkern-\thinmuskip\}}
\newcommand\mempty{\pmb{\emptyset}}
\newcommand\circled[1]{{\setlength\unitlength{1em}
  \begin{picture}(1,1)(0,0)
    \put(0.2,0.2){\circle{0.7}}
    \put(0,0){#1}
  \end{picture}}}
\newcommand\cq{{\smash{\circled{$\scriptstyle ?$}}}}

\begin{document}

\title[WSTS I: Completions]{Forward Analysis for WSTS, Part~{I:} Completions}

\author[LSV]{A. Finkel}{Alain Finkel}
\address[LSV]{LSV, ENS Cachan, CNRS; 61 avenue du pr\'esident Wilson,
  F-94230 Cachan}

\author[LSV,INRIA]{J. Goubault-Larrecq}{Jean Goubault-Larrecq}
\address[INRIA]{INRIA Saclay Ile-de-France}	
\email{{finkel,goubault}@lsv.ens-cachan.fr}  



\keywords{WSTS, forward analysis, completion, Karp-Miller procedure, domain theory, sober spaces, Noetherian spaces}


\begin{abstract}
  Well-structured transition systems provide the right foundation to
  compute a finite basis of the set of predecessors of the upward
  closure of a state.  The dual problem, to compute a finite
  representation of the set of successors of the downward closure of a
  state, is harder: Until now, the theoretical framework for
  manipulating downward-closed sets was missing.  We answer this
  problem, using insights from domain theory (dcpos and ideal
  completions), from topology (sobrifications), and shed new light on
  the notion of adequate domains of limits.
\end{abstract}

\maketitle


\section{Introduction}
\label{sec:intro}

The theory of well-structured transition systems (WSTS) is 20 years
old \cite{F90,FS:wsts,DBLP:journals/iandc/AbdullaCJT00}.  The most
often used result of this theory \cite{FS:wsts} is the backward
algorithm for computing a finite basis of the set $\uparrow
Pre^*(\uparrow s)$ of predecessors of the upward closure $\uparrow s$
of a state $s$.  The starting point of this paper is our desire to
compute $\downarrow Post^*(\downarrow s)$ in a similar way.  We then
need a theory to finitely (and effectively) represent downward-closed
sets, much as upward-closed subsets can be represented by their finite
sets of minimal elements.  This will serve as a basis for constructing
forward procedures.

The {\em cover\/}, ${\downarrow Post^*({\downarrow s})}$, contains
more information than the set of predecessors $\uparrow
Pre^*({\uparrow s})$ because it characterizes a good approximation of
the reachability set, while the set of predecessors describes the
states from which the system may fail; the cover may also allow the
computation of a finite-state abstraction of the system as a symbolic
graph.  Moreover, the backward algorithm needs a finite basis of the
upward closed set of bad states, and its implementation is, in
general, less efficient than a forward procedure: e.g., for lossy
channel systems, although the backward procedure always terminates,
only the non-terminating forward procedure is implemented in the tool
TREX \cite{ABJ:SRE}.

Except for some partial results
\cite{F90,DBLP:conf/lics/EmersonN98,GRvB:eec}, a general theory of
downward-closed sets is missing.  This may explain the scarcity of
forward algorithms for WSTS.  Quoting Abdulla {\em et al.\/}
\cite{DBLP:conf/formats/AbdullaDMN04}: ``Finally, we aim at developing
generic methods for building downward closed languages, in a similar
manner to the methods we have developed for building upward closed
languages in \cite{DBLP:journals/iandc/AbdullaCJT00}.  This would give
a general theory for forward analysis of infinite state systems, in
the same way the work in \cite{DBLP:journals/iandc/AbdullaCJT00} is
for backward analysis.''  Our contribution is to provide such a theory
of downward-closed sets.


\paragraph{\em Related Work.}

Karp and Miller \cite{KM:petri} proposed an algorithm that computes a
finite representation of the downward closure of the reachability set
of a Petri net.
Finkel \cite{F90} introduced the WSTS framework and generalized the
Karp-Miller procedure to a class of WSTS.  This is done by
constructing the completion of the set of states (by ideals, see
Section~\ref{sec:adl}) and in replacing the $\omega$-acceleration of
an increasing sequence of states (in Petri nets) by its least upper
bound (lub).  However, there are no effective finite representations
of downward closed sets in \cite{F90}.  Emerson and Namjoshi
\cite{DBLP:conf/lics/EmersonN98} considered a variant of WSTS (using
cpos,
but still without a theory of effective finite representations of
downward-closed subsets) for defining a Karp-Miller procedure to
broadcast protocols---termination is then not guaranteed
\cite{DBLP:conf/lics/EsparzaFM99}.
Abdulla {\em et al.\/} \cite{ABJ:SRE}
proposed a forward procedure for lossy channel systems using
downward-closed languages, coded as SREs.  Ganty, Geeraerts, and
others
\cite{GRvB:eec,DBLP:conf/vmcai/GantyRB06} proposed a forward procedure
for solving the coverability problem for WSTS equipped with an
effective adequate domain of limits.  This domain ensures that every
downward closed set has a finite representation; but no insight is
given how these domains can be found or constructed.  They applied
this to Petri nets and lossy channel systems.  Abdulla {\em et al.\/}
\cite{DBLP:conf/formats/AbdullaDMN04} proposed another symbolic
framework for dealing with downward closed sets for timed Petri nets.

We shall see that these constructions are special cases of our
completions (Section~\ref{sec:adl}).  We shall illustrate this in
Section~\ref{sec:concrete}, and generalize to a comprehensive
hierarchy of data types in Section~\ref{sec:domains}.  We briefly
touch the question of computing approximations of the cover in
Section~\ref{section-theory-for-forward-analysis-of-infinite-wsts},
although we shall postpone most of it to future work.  We conclude in
Section~\ref{sec:conc}.


\section{Preliminaries}
\label{sec:prelim}  

We shall borrow from theories of order, both from the theory of well
quasi-orderings, as used classically in well-structured transition
systems \cite{DBLP:journals/iandc/AbdullaCJT00,FS:wsts}, and from
domain theory \cite{AJ:domains,GHKLMS:contlatt}.  We should warn the
reader that this is one bulky section on preliminaries.  We invite her
to skip technical points first, returning to them on demand.

A {\em quasi-ordering\/} $\leq$ is a reflexive and transitive relation
on a set $X$.  It is a (partial) {\em ordering\/} iff it is
antisymmetric.  A set $X$ equipped with a partial ordering is a {\em
  poset\/}.

We write $\geq$ the converse quasi-ordering, $\approx$ the equivalence
relation $\leq \cap \geq$, $<$ associated strict ordering ($\leq
\setminus \approx$), and $>$ the converse ($\geq \setminus \approx$) of
$<$.  The {\em upward closure\/} $\uparrow E$ of a set $E$ is $\{y \in
X \mid \exists x \in E \cdot x \leq y\}$.  The {\em downward
  closure\/} $\downarrow E$ is $\{y \in X \mid \exists x \in E \cdot y
\leq x\}$.  A subset $E$ of $X$ is {\em upward closed\/} if and only
if $E = {\uparrow E}$, i.e., any element greater than or equal to some
element in $E$ is again in $E$.  {\em Downward closed\/} sets are
defined similarly.  When the ambient space $X$ is not clear from
context, we shall write $\downarrow_X E$, $\uparrow_X E$ instead of
$\downarrow E$, $\uparrow E$.

A quasi-ordering is {\em well-founded\/} iff it has no infinite
strictly descending chain, i.e., $x_0 > x_1 > \ldots > x_i > \ldots$.
An {\em antichain\/} is a set of pairwise incomparable elements.  A
quasi-ordering is {\em well\/} if and only it is well-founded and has
no infinite antichain.

There are a number of equivalent definitions for well quasi-orderings
(wqo).  One is that, from any infinite sequence $x_0, x_1, \ldots,
x_i, \ldots$, one can extract an infinite ascending chain $x_{i_0}
\leq x_{i_1} \leq \ldots \leq x_{i_k} \leq \ldots$, with $i_0 < i_1 <
\ldots < i_k < \ldots$.  Another one is that any upward closed subset
can be written $\uparrow E$, with $E$ {\em finite\/}.  Yet another,
topological definition \cite[Proposition~3.1]{Gou-lics07} is to say
that $X$, with its Alexandroff topology, is Noetherian.  The {\em
  Alexandroff topology\/} on $X$ is that whose opens are exactly the
upward closed subsets.  A subset $K$ is compact if it satisfies the
Heine-Borel property, i.e., every one may extract a finite subcover
from any open cover of $K$.  A topology is {\em Noetherian\/} iff every
open subset is compact, iff any increasing chain of opens stabilizes
\cite[Proposition~3.2]{Gou-lics07}.  We shall cite results from the
latter paper as the need evolves.

We shall be interested in rather particular topological spaces, whose
topology arises from order.  A {\em directed family\/} of $X$ is any
non-empty family ${(x_i)}_{i \in I}$ such that, for all $i,j \in I$,
there is a $k \in I$ with $x_i, x_j \leq x_k$.  The {\em Scott
  topology\/} on $X$ has as opens all upward closed subsets $U$ such
that every directed family ${(x_i)}_{i \in I}$ that has a least upper
bound $x$ in $X$ intersects $U$, i.e., $x_i \in U$ for some $i \in I$.
The Scott topology is coarser than the Alexandroff topology, i.e.,
every Scott-open is Alexandroff-open (upward closed); the converse
fails in general.  The Scott topology is particularly interesting on
{\em dcpos\/}, i.e., posets $X$ in which every directed family
${(x_i)}_{i \in I}$ has a least upper bound $\sup_{i \in I} x_i$.

The {\em way below\/} relation $\ll$ on a poset $X$ is defined by $x
\ll y$ iff, for every directed family ${(z_i)}_{i \in I}$ that has a
least upper bound $z \geq y$, then $z_i \geq x$ for some $i \in I$
already.  Note that $x \ll y$ implies $x \leq y$, and that $x' \leq x
\ll y \leq y'$ implies $x' \ll y'$.  However, $\ll$ is not reflexive
or irreflexive in general.  Write $\uuarrow E = \{y \in X \mid \exists
x \in E \cdot x \ll y\}$, $\ddarrow E = \{y \in X \mid \exists x \in E
\cdot y \ll x\}$.  $X$ is {\em continuous\/} iff, for every $x \in X$,
$\ddarrow x$ is a directed family, and has $x$ as least upper bound.
One may be more precise: A {\em basis\/} is a subset $B$ of $X$ such
that any element $x \in X$ is the least upper bound of a directed
family of elements way below $x$ {\em in $B$\/}.  Then $X$ is
continuous if and only if it has a basis, and in this case $X$ itself
is the largest basis.  In a continuous dcpo, $\uuarrow x$ is
Scott-open for all $x$, and every Scott-open set $U$ is a union of
such sets, viz.\ $U = \bigcup_{x \in U} {\uuarrow x}$
\cite{AJ:domains}.


$X$ is {\em algebraic\/} iff every element $x$ is the least upper
bound of the set of finite elements below $x$---an element $y$ is {\em
  finite\/} if and only if $y \ll y$.  Every algebraic poset is
continuous, and has a least basis, namely its set of finite elements.

$\nat$, with its natural ordering, is a wqo and an algebraic poset.
All its elements are finite, so $x \ll y$ iff $x \leq y$.  $\nat$ is
not a dcpo, since $\nat$ itself is a directed family without a least
upper bound.  Any finite product of continuous posets (resp.,
continuous dcpos) is again continuous, and the Scott-topology on the
product coincides with the product topology.  Any finite product of
wqos is a wqo.  In particular, $\nat^k$, for any integer $k$, is a wqo
and a continuous poset: this is the set of configurations of Petri
nets.

It is clear how to complete $\nat$ to make it a cpo: let $\nat_\omega$
be $\nat$ with a new element $\omega$ such that $n \leq \omega$ for
all $n \in \nat$.  Then $\nat_\omega$ is still a wqo, and a continuous
cpo, with $x \ll y$ if and only if $x \in \nat$ and $x \leq y$.  In
general, completing a wqo is necessary to extend coverability tree
techniques \cite{F90,GRvB:eec}.  Geeraerts {\em et al.\/} (op.\ cit.)
axiomatize the kind of completions they need in the form of so-called
{\em adequate domains of limits\/}.  We discuss them in
Section~\ref{sec:adl}.  For now, let us note that the second author
also proposed to use another notion of completion in another context,
known as {\em sobrification\/} \cite{Gou-lics07}.
We need to recap what this is about.

A topological space $X$ is always equipped with a {\em specialization
  quasi-ordering\/}, which we shall write $\leq$ again: $x \leq y$ if
and only if any open subset containing $x$ also contains $y$.  $X$ is
{\em $T_0$\/} if and only if $\leq$ is a partial ordering.  Given any
quasi-ordering $\leq$ on a set $X$, both the Alexandroff and the Scott
topologies admit $\leq$ as specialization quasi-ordering.  In fact,
the Alexandroff topology is the finest (the one with the most opens)
having this property.  The coarsest is called the {\em upper
  topology\/}; its opens are arbitrary unions of complements of sets
of the form $\downarrow E$, $E$ finite.  The latter sets $\downarrow
E$, with $E$ finite, will play an important role, and we call them the
{\em finitary closed\/} subsets.  Note that finitary closed subsets
are closed in the upper, Scott, and Alexandroff topologies, recalling
that a subset is {\em closed\/} iff its complement is open.  The {\em
  closure\/} $cl (A)$ of a subset $A$ of $X$ is the smallest closed
subset containing $A$.  A closed subset $F$ is {\em irreducible\/} if
and only if $F$ is non-empty, and whenever $F \subseteq F_1 \cup F_2$
with $F_1, F_2$ closed, then $F \subseteq F_1$ or $F \subseteq F_2$.
The finitary closed subset $\downarrow x = cl (\{x\})$ ($x \in X$) is
always irreducible.  A space $X$ is {\em sober\/} iff every
irreducible closed subset $F$ is the closure of a unique point, i.e.,
$F = {\downarrow x}$ for some unique $x$.  Any sober space is $T_0$,
and any continuous cpo is sober in its Scott topology.  Conversely,
given a $T_0$ space $X$, the space $\Sober (X)$ of all irreducible
closed subsets of $X$, equipped with upper topology of the inclusion
ordering $\subseteq$, is always sober, and the map $\eta_\Sober: x
\mapsto {\uparrow x}$ is a topological embedding of $X$ inside $\Sober
(X)$.  $\Sober (X)$ is the {\em sobrification\/} of $X$, and can be
thought as $X$ together with all missing limits from $X$.  Note in
particular that a sober space is always a cpo in its specialization
ordering \cite[Proposition~7.2.13]{AJ:domains}.

It is an enlightening exercise to check that $\Sober (\nat)$ is
$\nat_\omega$.  Also, the topology on $\Sober (\nat)$ (the upper
topology) coincides with that of $\nat_\omega$ (the Scott topology).
In general, $X$ is Noetherian if and only if $\Sober (X)$ is
Noetherian \cite[Proposition~6.2]{Gou-lics07}, however the upper and
Scott topologies do not always coincide \cite[Section~7]{Gou-lics07}.
In case of ambiguity, given any poset $X$, we write $X_a$ the space
$X$ with its Alexandroff topology.

Another important construction is the {\em Hoare powerdomain\/}
$\Hoare (X)$ of $X$, whose elements are the closed subsets of $X$,
ordered by inclusion.  (We do allow the empty set.)  We again equip it
with the corresponding upper topology.

\section{Completions of Wqos}
\label{sec:adl}

One of the central problems of our study is the definition of a {\em
  completion\/} of a wqo $X$, with all missing limits added.
Typically, the Karp-Miller construction \cite{KM:petri} works not with
$\nat^k$, but with $\nat_\omega^k$.  We examine several ways to
achieve this, and argue that they are the same, up to some details.

\paragraph{\em ADLs, WADLs.}  We start with Geeraerts {\em et al.\/}'s
axiomatization of so-called {\em adequate domain of limits\/} for
well-quasi-ordered sets $X$
\cite{GRvB:eec}.  No explicit constructions for such adequate domains
of limits is given, and they have to be found by trial and error.  Our
main result, below, is that there is a unique least adequate domain of
limits: the {\em sobrification\/} $\Sober (X_a)$ of $X_a$.  (Recall
that $X_a$ is $X$ with its Alexandroff topology.)  This not only gives
a concrete construction of such an adequate domain of limits, but also
shows that we do not have much freedom in defining one.

An {\em adequate domain of limits\/} \cite{GRvB:eec} (ADL) for a
well-ordered set $X$ is a triple $(L, \preceq, \gamma)$ where $L$ is a
set disjoint from $X$ (the set of {\em limits\/}); ($\Lim_1$) the map
$\gamma : L \cup X \to \pow (X)$ is such that $\gamma (z)$ is downward
closed for all $z \in L \cup X$, and $\gamma (x) = {\downarrow_X x}$
for all non-limit points $x \in X$; ($\Lim_2$) there is a limit point
$\top \in L$ such that $\gamma (\top) = X$; ($\Lim_3$) $z \preceq z'$
if and only if $\gamma (z) \subseteq \gamma (z')$; and ($\Lim_4$) for
any downward closed subset $D$ of $X$, there is a finite subset $E
\subseteq L \cup X$ such that $\widehat\gamma (E) = D$.  Here
$\widehat\gamma (E) = \bigcup_{z \in E} \gamma (z)$.

Requirement ($\Lim_2$) in \cite{GRvB:eec} only serves to ensure that
all closed subsets of $L \cup X$ can be represented as $\downarrow_{L
  \cup X} E$ for some finite subset $E$: the closed subset $L \cup X$
itself is then exactly $\downarrow_{L \cup X} \{\top\}$.  However,
($\Lim_2$) is unnecessary for this, since $L \cup X$ already equals
$\downarrow_{L \cup X} E$ by ($\Lim_3$), where $E$ is the finite
subset of $L \cup X$ such that $\widehat\gamma (E) = L \cup X$ as
ensured by ($\Lim_4$).  Accordingly, we drop requirement ($\Lim_2$):
\begin{definition}[WADL]
  \label{defn:adl}
  Let $X$ be a poset.  A {\em weak adequate domain of limits\/}
  (WADL) on $X$ is any triple $(L, \preceq, \gamma)$ satisfying
  ($\Lim_1$), ($\Lim_3$), and ($\Lim_4$).
\end{definition}

\begin{proposition}
  \label{thm:adl}
  Let $X$ be a poset.  Given a WADL $(L, \preceq, \gamma)$ on $X$,
  $\gamma$ defines an order-isomorphism from $(L \cup X, \preceq)$ to
  some subset of $\Hoare (X_a)$ containing $\Sober (X_a)$.

  Conversely, assume $X$ wqo, and let $Y$ be any subset of $\Hoare
  (X_a)$ containing $\Sober (X_a)$.  Then $(Y \setminus \eta_\Sober
  (X_a), \preceq, \gamma)$ is a weak adequate domain of limits, where
  $\gamma$ maps each $x \in X$ to $\downarrow_X x$ and each $F \in Y
  \setminus \eta_\Sober (X_a)$ to itself; $\preceq$ is defined by
  requirement ($\Lim_3$).
\end{proposition}
\proof
  The Alexandroff-closed subsets of $X$ are just its downward-closed
  subsets.  So $\gamma (z)$ is in $\Hoare (X_a)$ for all $z$, by
  ($\Lim_1$).  Let $Y$ be the image of $\gamma$.  By ($\Lim_3$),
  $\gamma$ defines an order-isomorphism of $L \cup X$ onto $Y$.  It
  remains to show that $Y$ must contain $\Sober (X_a)$.  Let $F$ be
  any irreducible closed subset of $X_a$.  By ($\Lim_4$), there is a
  finite subset $E \subseteq L \cup X$ such that $F = \bigcup_{x \in
    E} \gamma (x)$.  Since $F$ is irreducible, there must be a single
  $x \in E$ such that $F = \gamma (x)$.  So $F$ is in $Y$.

  Conversely, let $X$ be wqo, $L = Y \setminus \eta_\Sober (X_a)$, and
  $\gamma$, $\preceq$ be as in the Lemma.  Properties ($\Lim_1$) and
  ($\Lim_3$) hold by definition.  For ($\Lim_4$), note that $X_a$ is a
  Noetherian space, hence $\Sober (X_a)$ is, too
  \cite[Proposition~6.2]{Gou-lics07}.  However, by
  \cite[Corollary~6.5]{Gou-lics07}, every closed subset of a sober
  Noetherian space is finitary.  In particular, take any downward
  closed subset $D$ of $X$.  This is closed in $X_a$, hence its image
  $\eta_\Sober (D)$ by the topological embedding $\eta_\Sober$ is
  closed in $\eta_\Sober (X_a)$, i.e., is of the form $\eta_\Sober
  (X_a) \cap F$ for some closed subset $F$ of $\Sober (X_a)$.  Also,
  $D = \eta_\Sober^{-1} (F)$.  Since $\Sober (X_a)$ is both sober and
  Noetherian, $F$ is finitary, hence is the downward-closure
  $\downarrow_{\Sober (X)} E'$ of some finite subset $E'$ in $\Sober
  (X)$.  Let $E$ be the set consisting of the (limit) elements in $E'
  \cap L$, and of the (non-limit) elements $x \in X$ such that
  $\downarrow_X x \in E'$.  We obtain $\widehat\gamma (E) = \bigcup_{z
    \in E'} z$.  On the other hand, $D = \eta_\Sober^{-1} (F) = \{x
  \in X \mid {\downarrow x} \in \downarrow_{\Sober (X)} E'\} = \{x \in
  X \mid \exists z \in E' \cdot {\downarrow x} \subseteq z\} =
  \bigcup_{z \in E'} z = \widehat\gamma (E)$.  So ($\Lim_4$) holds.
  \qed

I.e., up to the coding
function $\gamma$, there is a unique {\em minimal\/} WADL on any given
wqo $X$: its sobrification $\Sober (X_a)$.  There is also a unique
largest one: its Hoare powerdomain $\Hoare (X_a)$.  An adequate domain
of limits in the sense of Geeraerts {\em et al.\/} \cite{GRvB:eec},
i.e., one that additionally satisfies ($\Lim_2$) is, up to
isomorphism, any subset of $\Hoare (X_a)$ containing $\Sober (X_a)$
plus the special closed set $X$ itself as top element.  We contend
that $\Sober (X_a)$ is, in general, the sole WADL worth considering.

\paragraph{\em Ideal completions.}

We have already argued that $\Sober (X)$, for any Noetherian space
$X$, was in a sense of completion of $X$, adding missing limits.
Another classical construction to add limits to some poset $X$ is its
{\em ideal completion\/} $Idl (X)$.  The elements of the ideal
completion of $X$ are its {\em ideals\/}, i.e., its downward-closed
directed families, ordered by inclusion.  $Idl (X)$ can be visualized
as a form of Cauchy completion of $X$: we add all missing limits of
directed families ${(x_i)}_{i \in I}$ from $X$, by declaring these
families to be their limits, equating two families when they have the
same downward-closure.  In $Idl (X)$, the finite elements are the
elements of $X$; formally, the map $\eta_{Idl} : X \to Idl (X)$ that
sends $x$ to $\downarrow x$ is an embedding, and the finite elements
of $Idl (X)$ are those of the form $\eta_{Idl} (x)$.  It turns out
that sobrification and ideal completion coincide, in a strong sense:
\begin{proposition}[\cite{Mislove:algebraic}]
  \label{prop:sober=idl}
  For any poset $X$, $\Sober (X_a) = Idl (X)$.
\end{proposition}
This is not just an isomorphism: the irreducible closed subsets of
$X_a$ are {\em exactly\/} the ideals.  Note also that $Idl (X)$ is
always an algebraic dcpo \cite[Proposition~2.2.22,
Item~4]{AJ:domains}.

When $X$ is wqo, any downward-closed subset of $X$ is a {\em finite\/}
union of ideals.  So $(Idl (X) \setminus X, {\subseteq}, \mathrm{id})$
is a WADL on $X$.  Proposition~\ref{thm:adl} and
Proposition~\ref{prop:sober=idl} entail this, and a bit more:
\begin{theorem}
  For any wqo $X$, $\Sober (X_a) = Idl (X)$ is the smallest WADL on $X$.
\end{theorem}

\paragraph{\em Well-based continuous cpos.}  There is a natural notion
of limit in dcpos: whenever ${(x_i)}_{i \in I}$ is a directed family,
consider $\sup_{i \in I} x_i$.  Starting from a wqo $X$, it is then
natural to look at some dcpo $Y$ that would contain $X$ as a basis.
In particular, $Y$ would be continuous.  This prompts us to define a
{\em well-based continuous dcpo\/} as one that has a well-ordered
basis---namely the original poset $X$.

This has several advantages.  First, in general there are several
notions of ``sets of limits'' of a given subset $A \subseteq Y$, but
we shall see that they all coincide in continuous posets.  Such sets
of limits are important, because these are what we would like
Karp-Miller-like procedures to compute, through acceleration
techniques.  Here are the possible notions.  First, define $\LUB_Y
(A)$ as the set of all least upper bounds in $Y$ of directed families
in $A$.  Second, $\IND_Y (A)$, the {\em inductive hull\/} of $A$ in
$Y$, is the smallest sub-dcpo of $Y$ containing $A$.  Finally, the
(Scott-topological) closure $cl (A)$ of $A$.
It is well-known that $cl (A)$ is the smallest {\em downward closed\/}
sub-dcpo of $Y$ containing $A$.  (Recall that any open is upward
closed, so that any closed set must be downward closed.)  In any dcpo
$Y$, one has $A \subseteq \LUB_Y (A) \subseteq \IND_Y (A) \subseteq cl
(A)$, and all inclusions are strict in general.  E.g., in $Y =
\nat_\omega$, take $A$ to be the set of even numbers.  Then $\LUB_Y
(A) = \IND_Y (A) = A \cup \{\omega\}$ while $cl (A) = \nat_\omega$.
While $\LUB_Y (A) = \IND_Y (A)$ in this case, there are cases where
$\LUB_Y (A)$ is itself not closed under least upper bounds of directed
families, and one has to iterate the $\LUB_Y$ operator to compute
$\IND_Y (A)$.  On continuous posets however, all these notions
coincide \cite[Appendix~A]{FGL:completions}.
\begin{proposition}
  \label{prop:IND=LUB}
  Let $Y$ be a continuous poset.  Then, for every downward-closed
  subset $A$ of $Y$, $\IND_Y (A) = \LUB_Y (A) = cl (A)$.
\end{proposition}
We shall use this in
Section~\ref{section-theory-for-forward-analysis-of-infinite-wsts}.
The key point now is that, again, well-based continuous dcpos coincide
with completions of the form $\Sober (X_a)$ or $Idl (X)$, and are
therefore WADLs \cite[Appendix~B]{FGL:completions}.  This even holds
for continuous dcpos having a well-founded (not well-ordered) basis:
\begin{proposition}
  \label{prop:wellbased:alg}
  Any continuous dcpo $Y$ with a well-founded basis is
  order-isomorphic to $Idl (X)$ for some well-ordered set $X$.  One
  may take the subset of finite elements of $X$ for $Y$.  If $Y$ is
  well-based, then $X$ is well-ordered.
\end{proposition}

\section{Some Concrete WADLs}
\label{sec:concrete}

We now build WADLs for several concrete posets $X$.  Following
Proposition~\ref{thm:adl}, it suffices to characterize $\Sober (X_a)$.
Although $\Sober (X_a) = Idl (X)$ (Proposition~\ref{prop:sober=idl}),
the mathematics of $\Sober (X_a)$ is easier to deal with than $Idl
(X)$.

\paragraph{\em $\nat^k$.}
We start with $X = \nat^k$, with the pointwise ordering.  We have
already recalled from \cite{Gou-lics07} that $\Sober (\nat^k_a)$ was,
up to isomorphism, ${(\nat_\omega)}^k$, ordered with the pointwise
ordering, where $\omega$ is a new element above any natural number.
This is the structure used in the standard Karp-Miller construction
for Petri nets \cite{KM:petri}.

\paragraph{\em $\Sigma^*$.}  Let $\Sigma$ be a finite alphabet.  The
{\em divisibility ordering\/} $|$ on $\Sigma^*$, a.k.a.\ the
subsequence (non-continuous subword) ordering, is defined by $a_1 a_2
\ldots a_n \mathrel{|} w_0 a_1 w_1 \allowbreak a_2 \ldots \allowbreak
a_n w_n$, for any letters $a_1, a_2, \ldots, a_n \in \Sigma$ and words
$w_0, w_1, \ldots, w_n \in \Sigma^*$.  There is a more general
definition, where letters themselves are quasi-well-ordered.  Our
definition is the special case where the wqo on letters is $=$, and is
the one required in verifying lossy channel systems \cite{AJ:lossy}.
Higman's Lemma states that $|$ is wqo on $\Sigma^*$.

Any upward closed subset $U$ of $\Sigma^*$ is then of the form
$\uparrow E$, with $E$ finite.  For any element $w = a_1 a_2 \ldots
a_n$ of $E$, $\uparrow w$ is the regular language $\Sigma^* a_1
\Sigma^* a_2 \Sigma^* \ldots \Sigma^* a_n \Sigma^*$.  Forward analysis
of lossy channel systems is instead based on simple regular
expressions (SREs).  Recall from \cite{ABJ:SRE} that an {\em atomic
  expression\/} is any regular expression of the form $a^?$, with $a
\in \Sigma$, or $A^*$, where $A$ is a non-empty subset of $\Sigma$.
When $A = \{a_1, \ldots, a_m\}$, we take $A^*$ to denote ${(a_1+
  \ldots + a_m)}^*$; $a^?$ denotes $\{a, \epsilon\}$.  A {\em
  product\/} is any regular expression of the form $e_1 e_2 \ldots
e_n$ ($n \in \nat$), where each $e_i$ is an atomic expression.  A {\em
  simple regular expression\/}, or {\em SRE\/}, is a sum, either
$\emptyset$ or $P_1 + \ldots + P_k$, where $P_1, \ldots, P_k$ are
products.  Sum is interpreted as union.  That SREs and products are
relevant here is no accident, as the following proposition shows.
\begin{proposition}
  \label{prop:Sigma*:SRE}
  The elements of $\Sober (\Sigma^*_a)$ are exactly the denotations of
  products.  The downward closed subsets of $\Sigma^*$ are exactly the
  denotations of SREs.
\end{proposition}
\proof
  The second part is well-known.  If $F = P_1 + \ldots + P_k$ is
  irreducible closed, then by irreducibility $k$ must equal $1$, hence
  $F$ is denoted by a product.  Conversely, it is easy to show that
  any product denotes an ideal, hence an element of $Idl (X) = \Sober
  (X_a)$ (Proposition~\ref{prop:sober=idl}).  \qed

Inclusion between products can then be checked in quadratic time
\cite{ABJ:SRE}.  Inclusion between SREs can be checked in polynomial
time, too, because of the remarkable property that $P_1 + \ldots + P_m
\subseteq P'_1 + \ldots + P'_n$ if and only if, for every $i$ ($1\leq
i\leq m$), there is a $j$ ($1\leq j\leq n$) with $P_i \subseteq P'_j$
\cite[Lemma~1]{ABJ:SRE}.Similar lemmas are given by Abdulla {\em et
  al.\/} \cite[Lemma~3, Lemma~4]{DBLP:conf/formats/AbdullaDMN04} for
more general notions of SREs on words on infinite alphabets, and for a
similar notion for finite multisets of elements from a finite set
(both will be special cases of our constructions of
Section~\ref{sec:domains}).  This is again no accident, and is a
general fact about Noetherian spaces:
\begin{proposition}
  \label{prop:Noeth:closed}
  Let $X$ be a Noetherian space, e.g., a wqo with its Alexandroff
  topology.  Every closed subset $F$ of $X$ is a finite union of
  irreducible closed subsets $C_1, \ldots, C_m$.  If $C'_1, \ldots,
  C'_n$ are also irreducible closed, Then $C_1 \cup \ldots \cup C_m
  \subseteq C'_1 \cup \ldots \cup C'_n$ if and only if for every $i$
  ($1\leq i\leq m$), there is a $j$ ($1\leq j\leq n$) with $C_i
  \subseteq C'_j$.
\end{proposition}
\proof For the first part, by the results of \cite{Gou-lics07},
$\Sober (X)$ is Noetherian and sober, which entails that $F$ can be
written $\downarrow \{x_1, \ldots, x_m\}$; now take $C_i =
\eta_\Sober^{-1} ({\downarrow x_i})$, $1\leq i\leq m$ (see
\cite[Appendix~C]{FGL:completions} for details).  The second part is
an easy consequence of irreducibility.  \qed

Proposition~\ref{prop:Noeth:closed} suggests to represent closed
subsets of $X$ as finite subsets $A$ of $\Sober (X)$, interpreted as
the closed set $\bigcup_{C \in A} C$.  When $X = \Sigma^*_a$, $A$ is a
finite set of products, i.e., an SRE.  When $X = \nat^k_a$, $A$ is a
finite subset of $\nat^k_\omega$, interpreted as ${\downarrow A} \cap
\nat^k$.

\paragraph{\em Finite Trees.}  All the examples given above are
well-known.  Here is one that is new, and also more involved than the
previous ones.  Let $\mathcal F$ be a finite signature of function
symbols with their arities.  We let $\mathcal F_k$ the set of function
symbols of arity $k$; $\mathcal F_0$ is the set of {\em constants\/},
and is assumed to be non-empty.
The set $\mathcal T (\mathcal F)$ is the set of ground terms built
from $\mathcal F$.  Kruskal's Tree Theorem states that this is
well-quasi-ordered by the {\em homeomorphic embedding\/} ordering
$\unlhd$, defined as the smallest relation such that, whenever $u = f
(u_1, \ldots, u_m)$ and $v = g (v_1, \ldots, v_n)$, $u \unlhd v$ if
and only if $u \unlhd v_j$ for some $j$, $1\leq j\leq n$, or $f=g$,
$m=n$, and $u_1 \unlhd v_1$, $u_2 \unlhd v_2$, \ldots, $u_m \unlhd
v_m$.  (As for $\Sigma^*$, we take a special case, where each function
has fixed arity.)

The structure of $\Sober (\mathcal T (\mathcal F)_a)$ is described
using an extension of SREs to the tree case.  This uses regular tree
expressions as defined in \cite[Section~2.2]{tata97}.  Let $\mathcal
K$ be a countably infinite set of additional constants, called {\em
  holes\/} $\Box$.  Most tree regular expressions are
self-explanatory, except Kleene star $L^{*,\Box}$ and concatenation $L
._\Box L'$.  The latter denotes the set of all terms obtained from a
term $t$ in $L$ by replacing all occurrences of $\Box$ by (possibly
different) terms from $L'$.  The language of a hole $\Box$ is just
$\{\Box\}$.  $L^{*,\Box}$ is the infinite union of the languages of
$\Box$, $L$, $L ._\Box L$, $L ._\Box L ._\Box L$, etc.
\begin{definition}[STRE]
  \label{defn:tree:SRE}
  {\em Tree products\/} and {\em product iterators\/} are defined
  inductively by:
  \begin{itemize}
  \item Every hole $\Box$ is a tree product.
  \item $f^? (P_1, \ldots, P_k)$ is a tree product, for any $f \in
    \Sigma_k$ and any tree products $P_1, \ldots, P_k$.  We take $f^?
    (P_1, \ldots, P_k)$ as an abbreviation for $f (P_1, \ldots, P_k) +
    P_1 + \ldots + P_k$.
  \item $(\sum_{i=1}^n C_i)^{*,\Box} ._\Box P$ is a tree product, for
    any tree product $P$, any $n\geq 1$, and any product iterators
    $C_i$ over $\Box$, $1\leq i\leq n$.  We write $\sum_{i=1}^n C_i$
    for $C_1 + C_2 + \ldots + C_n$.
  \item $f (P_1, \ldots, P_k)$ is a product iterator over $\Box$ for
    any $f \in \Sigma_k$, where: 1. each $P_i$, $1\leq i\leq k$ is
    either $\Box$ itself or a tree product such that $\Box$ is not in
    the language of $P_i$; and 2. $P_i = \Box$ for some $i$, $1\leq
    i\leq k$.
  \end{itemize}
  A {\em simple tree regular expression\/} (STRE) is a finite sum of
  tree products.
\end{definition}
A tree regular expression is {\em closed\/} iff it has no free hole,
where a hole is free in $f (L_1, \ldots, L_k)$, $L_1 + \ldots + L_k$,
or in $f^? (L_1, \ldots, L_k)$ iff it is free in some $L_i$, $1\leq
i\leq k$; the only free hole in $\Box$ is $\Box$ itself; the free
holes of $L^{*,\Box}$ are those of $L$, plus $\Box$; the free holes of
$L ._\Box L'$ are those of $L'$, plus those of $L$ except $\Box$.
E.g., $f^? (a^?, b^?)$ and $(f (\Box, g^? (a^?)) + f (g^? (b^?),
\Box))^{*,\Box} ._\Box f^? (a^?, b^?)$ are closed tree products.
Then \cite[Appendix~D]{FGL:completions}:
\begin{theorem}
  \label{thm:STRE}
  The elements of $\Sober (\mathcal T (\mathcal F)_a)$ are exactly the
  denotations of closed tree products.  The downward closed subsets of
  $\mathcal T (\mathcal F)$ are exactly the denotations of closed
  STREs.  Inclusion is decidable in polynomial time for tree products
  and for STREs.
\end{theorem}

\section{A Hierarchy of Data Types}
\label{sec:domains}

The sobrification WADL can be computed in a compositional way, as we
now show.  Consider the following grammar of data types of interest in
verification:\\
{\small
  $
  \begin{array}{rcll}
    D & ::= & \nat & \mbox{natural numbers} \\
    & \mid & A_\leq & \mbox{finite set $A$, quasi-ordered by $\leq$}\\
    & \mid & D_1 \times \ldots \times D_k & \mbox{finite product} \\
    & \mid & D_1 + \ldots + D_k & \mbox{finite, disjoint sum} \\
    & \mid & D^* & \mbox{finite words} \\
    & \mid & D^\circledast & \mbox{finite multisets}
  \end{array}
  $
}

By {\em compositional\/}, we mean that the sobrification of any data
type $D$ is computed in terms of the sobrifications of its arguments.
E.g., $\Sober (D^*_a)$ will be expressed as some extended form of
products over $\Sober (D_a)$.  The semantics of data types is the
intuitive one.  Finite products are quasi-ordered by the pointwise
quasi-ordering, finite disjoint sums by comparing elements in each
summand---elements from different summands are incomparable.  For any
poset $X$ (even infinite), $X^*$ is the set of finite words over $X$
ordered by the {\em embedding\/} quasi-ordering $\leq^*$: $w \leq^*
w'$ iff, writing $w$ as the sequence of $m$ letters $a_1 a_2 \ldots
a_m$, one can write $w'$ as $w_0 a'_1 w_1 a'_2 w_2 \ldots w_{m-1} a'_m
w'_m$ with $a_1 \leq a'_1$, $a_2 \leq a'_2$, \ldots, $a_m \leq a'_m$.
$X^\circledast$ is the set of finite multisets $\mopen x_1, \ldots,
x_n \mclose$ of elements of $X$, and is quasi-ordered by
$\leq^\circledast$, defined as: $\mopen x_1, \allowbreak x_2, \ldots,
\allowbreak x_m \mclose \leq^\circledast \mopen y_1, y_2, \ldots,
y_n\mclose$ iff there is an injective map $r : \{1, \ldots,
\allowbreak m\} \to \{1, \ldots, \allowbreak n\}$ such that $x_i \leq
y_{r (i)}$ for all $i$, $1\leq i\leq m$.
When $\leq$ is just equality, $m \leq^\circledast m'$ iff every
element of $m$ occurs at least as many times in $m'$ as in $m$: this
is the $\leq^m$ quasi-ordering considered, on finite sets $X$, by
Abdulla {\em et al.\/}
\cite[Section~2]{DBLP:conf/formats/AbdullaDMN04}.

The analogue of products and SREs for $D^*$ is given by the following
definition, which generalizes the $\Sigma^*$ case of
Section~\ref{sec:concrete}.  Note that $D$ is in general an {\em
  infinite\/} alphabet, as in \cite{DBLP:conf/formats/AbdullaDMN04}.
The following definition should be compared with \cite{ABJ:SRE}.  The
only meaningful difference is the replacement of $(a+\epsilon)$, where
$a$ is a letter, with $C^?$, where $C \in \Sober (X_a)$.  It should
also be compared with the {\em word language generators\/} of
\cite[Section~6]{DBLP:conf/formats/AbdullaDMN04}.  Indeed, the latter
are exactly our products on $A^\circledast$, where $A$ is a finite
alphabet (in our notation, $A_\leq$, with $\leq$ given as equality).
\begin{definition}[Product, SRE]
  \label{defn:H:SRE}
  Let $X$ be a topological space.  Let $X^*$ be the set of finite
  words on $X$.  For any $A, B \subseteq X^*$, let $A B$ be $\{ww'
  \mid w \in A, w' \in B\}$, $A^*$ be the set of words on $A$, $A^? =
  A \cup \{\epsilon\}$.

  {\em Atomic expressions\/} are either of the form $C^?$, with $C \in
  \Sober (X)$, or $A^*$, with $A$ a non-empty finite subset of $\Sober
  (X)$.  {\em Products\/} are finite sequences $e_1 e_2 \ldots e_k$,
  $k \in \nat$, and {\em SREs\/} are finite sums of products.  The
  denotation of atomic expressions is given by $\Eval {C^?} = C^?$,
  $\Eval {A^*} = (\bigcup_{C \in A} \Eval C)^*$; of products by $\Eval
  {e_1 e_2 \ldots e_k} = \Eval {e_1} \Eval {e_2} \ldots \Eval {e_k}$;
  of SREs by $\Eval {P_1 + \ldots + P_k} = \bigcup_{i=1}^k \Eval
  {P_i}$.

  Atomic expressions are ordered by $C^? \sqsubseteq {C'}^?$ iff $C
  \subseteq C'$; $C^?  \sqsubseteq {A'}^*$ iff $C \subseteq C'$ for
  some $C' \in A'$; $A^* \not\sqsubseteq {C'}^?$; $A^* \sqsubseteq
  {A'}^*$ iff for every $C \in A$, there is a $C' \in A'$ with $C
  \subseteq C'$.  Products are quasi-ordered by $e P \sqsubseteq e'
  P'$ iff (1) $e \not\sqsubseteq e'$ and $e P \sqsubseteq P'$, or (2)
  $e=C^?$, $e'={C'}^?$, $C \subseteq C'$ and $P \sqsubseteq P'$, or
  (3) $e'={A'}^*$, $e \sqsubseteq {A'}^*$ and $P \sqsubseteq e' P'$.
  We let $\equiv$ be $\sqsubseteq \cap \sqsupseteq$.
\end{definition}

\begin{definition}[$\circledast$-Product, $\circledast$-SRE]
  \label{defn:M:SRE}
  Let $X$ be a topological space.
  For any $A, B \subseteq X$, let $A \odot B = \{m \uplus m' \mid m
  \in A, m' \in B\}$, $A^\circledast$ be the set of multisets
  comprised of elements from $A$, $A^\cq = \{\mopen x\mclose \mid x
  \in A\} \cup \{\mempty\}$, where $\mempty$ is the empty multiset.

  The {\em $\circledast$-products\/} $P$ are the expressions of the
  form $A^\circledast \odot C_1^\cq \odot \ldots \odot C_n^\cq$, where
  $A$ is a finite subset of $\Sober (X)$, $n \in \nat$, and $C_1,
  \ldots, C_n \in \Sober (X)$.  Their denotation $\Eval P$ is
  $(\bigcup_{C \in A} C)^\circledast \odot \Eval {C_1}^\cq \odot
  \ldots \odot \Eval {C_n}^\cq$.  They are quasi-ordered by $P
  \sqsubseteq P'$, where $P = A^\circledast \odot C_1^\cq \odot
  C_2^\cq \odot \ldots \odot C_m^\cq$ and $P' = {A'}^\circledast \odot
  {C'_1}^\cq \odot {C'_2}^\cq \odot \ldots \odot {C'_n}^\cq$, iff: (1)
  for every $C \in A$, there is a $C' \in A'$ with $C \subseteq C'$,
  and (2) letting $I$ be the subset of those indices $i$, $1\leq i\leq
  m$, such that $C_i \subseteq C'$ for no $C' \in A'$, there is an
  injective map $r : I \to \{1, \ldots, n\}$ such that $C_i
  \subseteq C'_{r (i)}$ for all $i \in I$.  Let $\equiv$ be
  $\sqsubseteq \cap \sqsupseteq$.
\end{definition}

\begin{theorem}
  \label{thm:domains}
  For every data type $D$, $\Sober (D_a)$ is Noetherian, and is
  computed by: $\Sober (\nat_a) = \nat_\omega$; $\Sober ({A_\leq}_a) =
  A_\leq$; $\Sober ((D_1 \times \ldots \times D_k)_a) = \Sober
  ({D_1}_a) \times \ldots \times \Sober ({D_k}_a)$; $\Sober ((D_1 +
  \ldots + D_k)_a) = \Sober ({D_1}_a) + \ldots + \Sober ({D_k}_a)$;
  $\Sober (D^*)$ is the set of products on $D$ modulo $\equiv$,
  ordered by $\sqsubseteq$ (Definition~\ref{defn:H:SRE}); $\Sober
  (D^\circledast)$ is the set of $\circledast$-products on $D$ modulo
  $\equiv$, ordered by $\sqsubseteq$ (Definition~\ref{defn:M:SRE}).

  For any data type $D$, equality and ordering (inclusion) in $\Sober
  (D_a)$ is decidable in the polynomial hierarchy.
\end{theorem}
\proof
  We show that $\Sober (D_a)$ is Noetherian and is computed as given
  above, by induction on the construction of $D$.  We in fact prove
  the following two facts separately: (1) $\Sober (D)$ is Noetherian
  ($D$, not $D_a$), where $D$ is topologized in a suitable way, and
  (2) $D=D_a$.

  To show (1), we topologize $\nat$ and $A_\leq$ with their
  Alexandroff topologies, sums and products with the sum and product
  topologies respectively; $X^*$ with the {\em subword topology\/},
  viz.\ the smallest containing the open subsets $X^* U_1 X^* U_2 X^*
  \ldots X^* U_n X^*$, $n \in \nat$, $U_1$, $U_2$, \ldots, $U_n$ open
  in $X$; and $X^\circledast$ with the {\em sub-multiset topology\/},
  namely the smallest containing the subsets $X^\circledast \odot U_1
  \odot U_2 \odot \ldots \odot U_n$, $n \in \nat$, where $U_1$, $U_2$,
  \ldots, $U_n$ are open subsets of $X$.  The case of $\nat$ has
  already been discussed above.  When $A_\leq$ is finite, it is both
  Noetherian and sober.  The case of finite products is by
  \cite[Section~6]{Gou-lics07}, that of finite sums by
  \cite[Section~4]{Gou-lics07}.  The cases of $X^*$, resp.\
  $X^\circledast$, are dealt with in \cite[Appendices~E,
  F]{FGL:completions}.

  To show (2), we appeal to a series of coincidence lemmas, showing
  that ${(X^*)}_a = X_a^*$ and that ${(X^\circledast)}_a =
  X_a^\circledast$ notably.  The other cases are obvious.

  Finally, we show that inclusion and equality are decidable in the
  polynomial hierarchy.  For this, we show in the appendices that
  inclusion on $\Sober (D^*)$ is $\sqsubseteq$ on products, and is
  decidable by a polynomial time algorithm modulo calls to an oracle
  deciding inclusion in $\Sober (D)$.  This is by dynamic programming.
  Inclusion in $\Sober (D^\circledast)$ is $\sqsubseteq$ on
  $\circledast$-products, and is decidable by a non-deterministic
  polynomial time algorithm modulo a similar oracle.  We conclude
  since the orderings on $\nat_\omega$ and on $A_\leq$ are
  polynomial-time decidable, while inclusion in $\Sober (D_1 \times
  \ldots \times D_k) \cong \Sober (D_1) \times \ldots \times \Sober
  (D_k)$ and in $\Sober (D_1 + \ldots + D_k) \cong \Sober (D_1) +
  \ldots + \Sober (D_k)$ are polynomial time modulo oracles deciding
  inclusion in $\Sober (D_i)$, $1\leq i\leq k$.  \qed

Look at some special cases of this construction.  First, $\nat^k$ is
the data type $\nat \times \ldots \times \nat$, and we retrieve that
$\Sober (\nat^k) = \nat_\omega^k$.  Second, when $A$ is a finite
alphabet, $A^*$ is given by products, as given in the $\Sigma^*$
paragraph of Section~\ref{sec:concrete}; i.e., we retrieve the
products (and SREs) of Abdulla {\em et al.\/} \cite{ABJ:SRE}.  The
more complicated case $(A^\circledast)^*$ was dealt with by Abdulla
{\em et al.\/} \cite{DBLP:conf/formats/AbdullaDMN04}.  We note that
the elements of $\Sober ((A^\circledast)^*_a)$ are exactly their {\em
  word language generators\/}, which we retrieve here in a principled
way.  Additionally, we can deal with more complex data structures such
as, e.g., $(((\nat \times A_\leq)^* \times
\nat)^\circledast)^\circledast$.

Finally, note that (1) and (2) are two separate concerns in the proof
of Theorem~\ref{thm:domains}.  If we are ready to relinquish orderings
for the more general topological route, as advocated in
\cite{Gou-lics07}, we could also enrich our grammar of data types with
infinite constructions such as $\pow (D)$, where $\pow (D)$ is
interpreted as the powerset of $D$ with the so-called lower Vietoris
topology.  In fact, $\Sober (\pow (X)) \cong \Hoare (X)$ is Noetherian
whenever $X$ is, and its elements can be represented as {\em finite\/}
subsets $A$ of $\Sober (X)$, interpreted as $\bigcup_{C \in A} C$
\cite[Appendix~G]{FGL:completions}.  In a sense, while $\Sober (X_a) =
Idl (X)$ for all ordered spaces $X$, the sobrification construction is
more robust than the ideal completion.

\section{Completing WSTS, or: Towards Forward Procedures Computing the Cover}
\label{section-theory-for-forward-analysis-of-infinite-wsts}

We show how one may use our completions on wqos to deal with forward
analysis of well-structured systems.  We shall describe this in more
detail in another paper.  First note that any data type $D$ of
Section~\ref{sec:domains} is suited to applying the expand, enlarge
and check algorithm \cite{GRvB:eec} out of the box to this end, since
then $\Sober (D_a)$ is (the least) WADL for $D$.  We instead explore
extensions of the Karp-Miller procedure \cite{KM:petri}, in the spirit
of Finkel \cite{F90} or Emerson and Namjoshi
\cite{DBLP:conf/lics/EmersonN98}.  While the latter assumes an already
built completion, we construct it.  Also, we make explicit how this
kind of acceleration-based procedure really computes the cover, i.e.,
$\downarrow Post^*({\downarrow x})$, in Proposition~\ref{prop:Post}.

Recall that a {\em well-structured transition system\/} (WSTS) is a
triple $S = (X, \leq, (\delta_i)_{i=1}^n)$, where $X$ is
well-quasi-ordered by $\leq$, and each $\delta_i : X \to X$ is a
partial monotonic transition function.  (By ``partial monotonic'' we
mean that the domain of $\delta_i$ is upward closed, and $\delta_i$ is
monotonic on its domain.)  Letting $Pre (A) = \bigcup_{i=1}^n
\delta_i^{-1} (A)$, $Pre^0 (A) = A$, and $Pre^* (A) = \bigcup_{k \in
  \nat} Pre^k (A)$, it is well-known that any upward closed subset of
$X$ is of the form $\uparrow E$ for some finite $E \subseteq X$, and
that $Pre^* ({\uparrow E})$ is an upward-closed subset ${\uparrow
  E'}$, $E'$ finite, that arises as $\bigcup_{k=0}^m Pre^k ({\uparrow
  E})$ for some $m \in \nat$.  Hence, provided $\leq$ is decidable and
$\delta_i^{-1} ({\uparrow E})$ is computable for each finite $E$, it
is decidable whether $x \in Pre^* ({\uparrow E})$, i.e., whether one may reach
${\uparrow E}$ from $x$ in finitely many steps.  It is equivalent to check
whether $y \in {\downarrow Post^* ({\downarrow x})}$ for some $y \in
E$, where $Post (A) = \bigcup_{i=1}^n \delta_i (A)$, $Post^0 (A) = A$,
and $Post^* (A) = \bigcup_{k \in \nat} Post^k (A)$.

All the existing symbolic procedures that attempt to compute
$\downarrow Post^*({\downarrow x})$, even with a finite number of
accelerations (e.g., Fast, Trex, Lash), can only compute subsets of
the larger set $\LUB ({\downarrow Post^*({\downarrow x})})$.  In
general, $\LUB ({\downarrow Post^*({\downarrow x})})$ does not admit a
finite representation.  On the other hand, we know that the
Scott-closure $cl (Post^* ({\downarrow x}))$, as a closed subset of
$Idl (X)$ (intersected with $X$ itself), is always finitary.  Indeed,
it is also a closed subset of $\Sober (X_a)$
(Proposition~\ref{prop:sober=idl}), which is represented as the
downward closure of finitely many elements of $\Sober (X_a)$.  Since
$Y = Idl (X)$ is continuous, Proposition~\ref{prop:IND=LUB} allows us
to conclude that $\LUB_Y ({\downarrow Post^*({\downarrow x})}) = cl
(Post^* ({\downarrow x}))$ is finitary---hence representable provided
$X$ is one of the data types of Section~\ref{sec:domains}.

This leads to the following construction.  Any partial monotonic map
$f : X \to Y$ between quasi-ordered sets lifts to a {\em continuous\/}
partial map $\Sober f : \Sober (X_a) \to \Sober (Y_a)$: for each
irreducible closed subset (a.k.a., ideal) $C$ of $\Sober (X_a)$,
either $C \cap \dom f \neq \emptyset$ and $\Sober f (C) = {\downarrow
  f (C)} = \{y \in Y \mid \exists x \in C \cap \dom f \cdot y \leq f
(x)\}$, or $C \cap \dom f = \emptyset$ and $\Sober f (C)$ is
undefined.  The {\em completion\/} of a WSTS $S = (X, \leq,
(\delta_i)_{i=1}^n)$ is then the transition system $\widehat S =
(\Sober (X_a), \subseteq, (\Sober \delta_i)_{i=1}^n)$.

For example, when $X = \nat^k$, and $S$ is a Petri net with
transitions $\delta_i$ defined as $\delta_i (\vec x) = \vec x + \vec
d_i$ (where $\vec d_i \in \Z^k$; this is defined whenever $\vec x+\vec
d \in \nat^k$), then $\widehat S$ is the transition system whose set
of states is $\Sober (X) = \nat_\omega^k$, and whose transition
functions are: $\Sober \delta_i (\vec x) = \vec x + \vec d_i$,
whenever this has only non-negative coordinates, taking the convention
that $\omega+d = \omega$ for any $d \in \Z$.

We may emulate lossy channel systems through the following {\em
  functional-lossy\/} channel systems (FLCS).
For simplicity, we assume just one channel and no local state; the
general case would only make the presentation more obscure.  An FLCS
differs from an LCS in that it loses only the least amount of messages
needed to enable transitions.  Take $X = \Sigma^*$ for some finite
alphabet $\Sigma$ of messages; the transitions are either of the form
$\delta_i (w) = w a_i$ for some fixed letter $a_i$ (sending $a_i$ onto
the channel), or of the form $\delta_i (w) = w_2$ whenever $w$ is of
the form $w_1 a_i w_2$, with $w_1$ not containing $a_i$ (expecting to
receive $a_i$).  Any LCS is cover-equivalent to the FLCS with the same
sends and receives, where two systems are {\em cover-equivalent\/} if
and only if they have the same sets $\downarrow {Post^* (F)}$ for any
downward-closed $F$.  Equating $\Sober (\Sigma^*_a)$ with the set of
products, as advocated in Section~\ref{sec:concrete}, we find that
transition functions of the first kind lift to $\Sober \delta_i (P) =
P a_i^?$ , while transition functions of the second kind lift to:
$\Sober \delta_i (\epsilon)$ is undefined, $\Sober \delta_i (a^? P) =
\Sober \delta_i (P)$ if $a_i\neq a$, $\Sober \delta_i (a_i^? P) = P$,
$\Sober \delta_i (A^* P) = \Sober \delta_i (P)$ if $a_i \not\in A$,
$\Sober \delta_i (A^* P) = A^* P$ otherwise.  This is exactly how Trex
computes successors \cite[Lemma~6]{ABJ:SRE}.

In general, the results of Section~\ref{sec:domains} allow us to use
any domain of datatypes $D$ for the state space $X$ of $S$.  The
construction $\widehat S$ then generalizes all previous constructions,
which used to be defined specifically for each datatype.

The Karp-Miller algorithm in Petri nets, or the Trex procedure for
lossy channel systems, gives information about the cover $\downarrow
Post^*({\downarrow x})$.  This is true of {\em any\/} completion
$\widehat S$ as constructed above:
\begin{proposition}
  \label{prop:Post}
  Let $S$ be a WSTS.  Let $\widehat{Post}$ be the $Post$ map of the
  completion $\widehat S$.  For any closed subset $F$ of $\Sober
  (X_a)$, $\widehat{Post} (F) = cl (Post (F \cap X))$, and
  $\widehat{Post}^* (F) = cl (Post^* (F \cap X))$.  Hence, for any
  downward closed subset $F$ of $X$, ${\downarrow {Post (F)}} = X \cap
  \widehat{Post} (F)$, ${\downarrow {Post^* (F)}} = X \cap
  \widehat{Post}^* (F)$.
\end{proposition}
\proof Let $F$ be closed in $\Sober (X_a)$. $\widehat{Post} (F) =
\bigcup_{i=1}^n cl (\delta_i (F)) = cl (\bigcup_{i=1}^n \delta_i (F))
= cl (Post (F))$, since closure commutes with (arbitrary) unions.  We
then claim that $\widehat{Post}^k (F) = cl (Post^k (F))$ for each $k
\in \nat$.  This is by induction on $k$.  The cases $k=0,1$ are
obvious.  When $k\geq 2$, we use the fact that, for any continuous
partial map $f$: $(*)$ $cl (f (cl (A))) = cl (f (A))$.  Then
$\widehat{Post}^k (F) = \bigcup_{i=1}^n cl (\delta_i
(\widehat{Post}^{k-1} (F))) = \bigcup_{i=1}^n cl (\delta_i (cl
(Post^{k-1} (F)))) = \bigcup_{i=1}^n cl (\delta_i (Post^{k-1} (F)))$
(by $(*)$) $= cl (Post^k (F))$.  Finally, $\widehat{Post}^* (F) =
\bigcup_{k \in \nat} \widehat{Post}^k (F) = \bigcup_{k \in \nat} cl
(Post^k (F)) = cl (Post^* (F))$.  We conclude, since for any $A
\subseteq X$, $\downarrow A$ is the closure of $A$ in $X_a$; the
topology of $X_a$ is the subspace topology of that of $\Sober (X_a)$;
so, writing $cl$ for closure in $\Sober (X_a)$, $\downarrow A = X \cap
cl (A)$.  \qed

Writing $F$ as the finite union $C_1 \cup \ldots \cup C_k$, where
$C_1, \ldots, C_k \in \Sober (X_a)$, $\widehat{Post} (F)$ is
computable as $\bigcup_{1\leq i_1, \ldots, i_n \leq k} \Sober \delta_1
(C_{i_1}) \cup \ldots \cup \Sober \delta_n (C_{i_n})$, assuming
$\Sober \delta_i$ computable for each $i$.  (We take $\Sober \delta_j
(C_i)$ to mean $\emptyset$ if undefined, for notational convenience.)
Although $\Sober \delta_i$ may be uncomputable even when $\delta_i$
is, it is computable on most WSTS in use.  This holds, for example,
for Petri nets and lossy channel systems, as exemplified above.

So it is easy to compute $\downarrow {Post ({\downarrow x})}$, as (the
intersection of $X$ with) $\widehat{Post} ({\downarrow x})$.
Computing $\downarrow {Post^* ({\downarrow x})}$ (our goal) is also
easily computed as $\widehat{Post}^* ({\downarrow x})$ (intersected
with $X$ again), using acceleration techniques for loops.  This is
what the Karp-Miller construction does for Petri nets, what Trex does
for lossy channel systems \cite{ABJ:SRE}.  (We examine termination
issues below.)
Our framework generalizes all these procedures, using a weak
acceleration assumption, whereby we assume that we can compute the
least upper bound of the values of loops iterated $k$ times, $k \in
\nat$.  For any continuous partial map $g : Y \to Y$ (with open
domain) on a dcpo $Y$, let the {\em iteration\/} $\overline g$ be the
map of domain $\dom g$ such that $\overline g (y)$ is the least upper
bound of ${(g^k (y))}_{k \in \nat}$ if $y < g (y)$, and $g (y)$
otherwise.  Let $\Delta = \{\Sober \delta_1, \ldots, \Sober
\delta_n\}$, $\Delta^*$ be the set of all composites of finitely many
maps from $\Delta$.  Our {\em acceleration assumption\/} is that one
can compute $\overline g (y)$ for any $g \in \Delta^*$, $y \in \Sober
(X_a)$.  The following procedure then computes $\downarrow {Post^*
  ({\downarrow x})}$, as (the intersection of $X$ with)
$\widehat{Post}^* ({\downarrow x})$, itself represented as a finite
union of elements of $\Sober (X_a)$: initially, let $A$ be $\{x\}$;
then, while $\widehat{Post} (A) \not\subseteq {\downarrow A}$, choose
fairly $(g, a) \in \Delta^* \times A$ such that $a \in \dom g$ and add
$\overline g (a)$ to $A$.  If this terminates, $A$ is a finite set
whose downward closure is exactly $\downarrow {Post^* ({\downarrow
    x})}$.  Despite its simplicity, this is the essence of the
Karp-Miller procedure, generalized to a large class of spaces $X$.

Termination is ensured for flat systems, i.e., systems whose control
graph has no nested loop, as one only has to compute the effect of a
finite number of loops.  In general, the procedure terminates on {\em
  cover-flattable\/} systems, that is systems that are
cover-equivalent to some flat system.  Petri nets are cover-flattable,
while, e.g., not all LCS are: recall that, in an LCS, $\downarrow
{Post^* ({\downarrow x})}$ is {\em always\/} representable as an SRE,
however not effectively so.



\section{Conclusion and Perspectives}
\label{sec:conc}

We have developed the first comprehensive theory of downward-closed
subsets, as required for a general understanding of forward analysis
techniques of WSTS.  This generalizes previous domain proposals on
tuples of natural numbers, on words, on multisets, allowing for nested
datatypes, and infinite alphabets.  Each of these domains is
effective, in the sense that each has finite presentations with a
decidable ordering.  We have also shown how the notion of
sobrification $\Sober (X_a)$ was in a sense inevitable
(Section~\ref{sec:adl}), and described how this applied to compute
downward closures of reachable sets of configurations in WSTS
(Section~\ref{section-theory-for-forward-analysis-of-infinite-wsts}).
We plan to describe such new forward analysis algorithms, in more
detail, in papers to come.

\bibliographystyle{abbrv}
\bibliography{post}

\end{document}